\documentstyle[preprint,aps,prl,epsf]{revtex}
\begin{document}
\title{Brownian motion at absolute zero}
\author{Supurna Sinha}
\address{Department of Physics, Syracuse University, 
Syracuse, New York 1,3244-1130}
\author{Rafael D. Sorkin}
\address{Enrico Fermi Institute, 5640 S. Ellis Avenue, Chicago, 
Illinois, 60637$^{*}$}
\maketitle
\widetext

\begin{abstract}
We derive a general quantum formula giving the mean-square displacement of a 
diffusing particle as a function of time. Near {\bf 0 K} we find a 
universal logarithmic behavior (valid for times longer than the relaxation 
time), and deviations from classical behavior can also be significant at 
larger values of time and temperature. Our derivation depends neither 
on the specific composition of the heat bath nor on the strength of the 
coupling between the bath and the particle. An experimental regime of 
microseconds and microdegrees Kelvin would elicit the pure logarithmic 
diffusion. 
\end{abstract}

\noindent
The so-called fluctuation-dissipation theorem--which relates the thermal 
fluctuations of a variable $x$ to the response of that variable to a weak 
external force--is usually described as generalizing the Smoluchowski- 
Einstein relation for Brownian motion, $D = kT\mu$; but it is not easy 
to find in the literature any explicit derivation of this relation as 
a direct corollary of the theorem. In this paper we will provide such a 
derivation under the assumption that the times involved are long compared 
to the relaxation time $\tau$, as defined below. But, because the 
fluctuation-dissipation theorem is really a quantum-mechanical 
relationship, it will tell us something more than just the laws of 
classical diffusion, which will emerge only in the limit 
$\hbar \rightarrow O$, or equivalently in the limit of long times and 
high temperatures. In the opposite limit where 
$kT\Delta t \ll \hbar$, the usual linear dependence 
$\Delta x^{2} \sim \Delta t$, will turn out to give 
way to a universal behavior $\Delta x^{2} \sim \ln \Delta t$, which 
probably should be interpreted as a 
diffusion driven by quantal zero-point motions rather than by thermal 
kinetic energy. The logarithmic behavior will follow from a general formula 
(12) for $\langle \Delta x^{2} \rangle$, which will hold for all 
times long compared to $\tau$, given the 
assumption of constant mobility $\mu$. In what follows we will derive this 
general formula, discuss the limiting cases just alluded to, and show that 
some deviations from classical behavior may be observable on the basis of 
current experimental technique.\\

\noindent
In recent years there have been several efforts$^{1-4}$ to understand the 
dynamics of a quantum particle coupled to a heat bath. Insofar as our work 
overlaps those efforts, our results appear to agree. The main difference is 
that the cited papers make far-reaching assumptions about the nature of 
the medium (heat bath) in which the particle moves, and require the coupling 
between particle and bath to be linear (meaning in effect that the coupling 
is weak). In contrast, we only use that the {\it response} to a 
weak {\it external} perturbation is 
linear, allowing the coupling of the particle to the bath and/or 
environment itself to be strong, as it will in fact be in most situations. 
On the other hand we will predict only the mean-square displacement, 
whereas the more special treatments can in principle yield the full density 
operator as a function of $\Delta t$. 

\begin{center}
{\bf THE FLUCTUATION-DISSIPATION\protect\\*[0.1cm]
THEOREM IN THE TIME DOMAIN}\\
\end{center}

\noindent
The fluctuation-dissipation theorem, as usually stated, refers to 
the Fourier transforms of the autocorrelation and response functions. 
Let $x(t)$ be some dynamical variable (operator) in the Heisenberg 
picture, and let $f(t)$ be an infinitely weak external force applied 
to $x$ at time $t$. (We will not need the more general form of the 
theorem in which the external coupling is to a different variable $y$.) 
the response function $R(t)$ is defined by the relation 

\begin{equation}
\langle x(t) \rangle_{f} - \langle x \rangle_{0} = \int {\bf R}(t - s)f(s)ds
\end{equation}
where $\langle \cdots \rangle_{f}$ denotes the expectation value in the 
presence of the force, assuming the system of which $x$ is a variable 
to have been in thermal equilibrium with temperature $T$ at early 
times; and $\langle \cdots \rangle_{0}$ is the same 
expectation value for zero force. Also let 
\begin{eqnarray*}
C(t) = \frac{1}{2} \langle x(t)x(0) + x(0)x(t)\rangle
\end{eqnarray*}
be the ''autocorrelation" or ''two-point" function in equilibrium at 
temperature $T$. [Or, if you prefer, you can subtract off 
$\langle x(t) \rangle \langle x(0) \rangle = \langle x(0) \rangle^{2}$ from 
this definition without invalidating what follows. This 
would be equivalent to working with $x - \langle x \rangle$ in place of 
$x$.] Then the fluctuation-dissipation theorems stated in the frequency 
domain is (with $\beta = 1/kT)$
\begin{equation} 
{\bf Im}\tilde{R}(\nu) = \hbar^{-1}\tanh (\pi\beta\hbar \nu)\sim{C}(\nu).
\end{equation}  
[We are using the following definition of fourier transform 
$\cal{F} \equiv ( \cdots )$: 
\begin{eqnarray*}
\tilde{\phi}(\nu) = \int dt 1^{\nu t}\phi^{*}(t),
\end{eqnarray*}
where $1^{x} \equiv e^{2\pi ix}.]$ \\

\noindent
Our first job is to transform this relation to the time domain. To that end, 
let us introduce in place of $R(t)$ (which vanishes for $t < 0$  by 
virtue of causality) the equivalent odd function 
\begin{eqnarray*}
\breve{R}(t) = {\rm sgn}(t)R(|t|).
\end{eqnarray*}
It is then easy to check that $2i{\rm Im}{\cal F}(\breve{R}) = {\cal F}
(\breve{R})$, whence (2) can be
written in the equivalent form:
\begin{equation}
{\cal F}(\breve{R}) = \frac{2i}{\hbar}\tanh (\pi\beta\hbar \nu){\cal 
F}(C).
\end{equation}

\noindent
[In fact it is actually this form, rather than (2), that comes out initially 
in the most straightforward derivation of the fluctuation-dissipation 
theorem; it is thus more appropriate to view (2) as a consequence of (3) 
rather than vice versa.] By taking the Fourier transform of (3) we could now 
express $R(t)$ as a convolution of C $(t)$, but our main interest here is to do the 
opposite. Let us therefore solve (3) for C, obtaining 

\begin{equation}
\bar{C}(\nu) = (- i\hbar / 2) \coth (\pi\beta\hbar \nu) [{\cal 
F}(\breve{R})] (\nu) + c\delta (\nu),
\end{equation}
where $c$ is a constant and where, for definiteness, the principal 
part of coth may be taken: $P(\coth x) = d/dx \ln \sinh |x|$. The 
ambiguity in $1/\tanh (\pi\beta\hbar \nu)$ is just a term proportional to 
$\delta (\nu)$, which would drop out of (4) anyway, since it would be 
multiplying the odd function ${\cal F}(\breve{R})$.] The Fourier 
transform of (4) reads 
\begin{equation}
C = \frac{i\hbar}{2} {\cal F} (\coth \pi\beta\hbar \nu) *\breve{R} + c ,
\end{equation}
determining $C$, up to an additive constant, in terms of the Fourier transform 
\begin{equation}
{\cal F}(\coth\pi\beta\hbar \nu) = (i / \beta\hbar) \coth (\pi t / 
\beta\hbar).
\end{equation}
In Eq. (6), the coth on the right-hand side is also to be understood as a 
principal part, but unlike before, this choice is forced on us, because the 
addition of any $\delta (t)$ piece to $\coth (\pi t / \beta\hbar)$ 
would spoil its 
oddness, in disagreement with the oddness of the left-hand side of (6). 
Understanding all coth's to be principal parts, then, we have finally 
(in view also of the definition of $\tilde{R}$) the following explicit 
formula for $C(t)$ in terms of $R(t)$:
\begin{eqnarray}
C(t) = \frac{1}{2\beta} \int^{\infty}_{-\infty}&&dt^{\prime}
{\rm sgn}(t^{\prime} - t) R(|t^{\prime} - t|)\nonumber\\
&&\times\coth (\pi t^{\prime} / \beta\hbar) + c .
\end{eqnarray}
[The appearance of the undetermined constant $c$ is due to the possibility of 
redefining the zero of $x$ without affecting (1). By working with the 
alternative definition of $C(t)$ mentioned just before Eq. (2), 
we would remove this ambiguity, and correspondingly could set $c =0$, given 
some assumptions on the asymptotic behavior of $x$ and $R$.]\\

\begin{center}
{\bf THE MEAN-SQUARE DISPLACEMENT $\langle \Delta x^{2} \rangle$} \\
\end{center}

\noindent
Now the mean-square displacement of $x$ due to equilibrium fluctuations 
in 
time $\Delta t$ is $\langle \Delta x^{2} \rangle$, where 
$\Delta x \equiv x (t + \Delta t) - x(t)$. 
Taking $t = 0$ for convenience, we have (since the 
equilibrium state is time independent) 
\begin{eqnarray}
\langle \Delta x^{2} \rangle = \langle[x(\Delta t)&-& x(0)]^{2} \rangle 
= \langle x(\Delta t)^{2} \rangle + \langle x(0)^{2} \rangle \nonumber\\
&-&\langle\{x(\Delta t), x(0)\} \rangle = 2C(0) - 2C(\Delta t),\nonumber\\
{\rm or}\hspace*{4cm}\frac{1}{2} \langle \Delta x^{2} \rangle = C(0) - C(\Delta t).
\end{eqnarray}
Combining this result with (7) gives us a general equation for 
$\langle \Delta x^{2} \rangle$ in terms of the response function $R$:
\begin{eqnarray}
\frac{1}{2} \langle \Delta x^{2} \rangle = \frac{1}{2\beta} \int^{\infty}_{0}
dt^{\prime} R(t^{\prime})&&[2 \coth \Omega t^{\prime} - \coth\Omega (t^{\prime} + t)\nonumber\\
&& -\coth\Omega (t^{\prime} -t)],
\end{eqnarray}
where for brevity we have set $\Omega = \pi / \beta\hbar$. Here, as before, 
the principal part of the coth is to be understood. Notice that the 
undetermined constant $c$ in (7) has dropped out of this result. \\

\begin{center}
{\bf QUANTUM BROWNIAN MOTION}\\
\end{center}

\noindent
At this stage, let us specialize $x$ to be a Cartesian coordinate of an 
otherwise free particle immersed in a homogeneous medium with temperature 
$T$. For an idealized inertialess Brownian particle, the response to a weak 
external force would be immediate motion at velocity $v = \mu f,\;\;\mu$ being 
the``mobility;" in other words, $R$ would be the step function 
$R(t) = \mu\Theta(t)$. However this idealization is 
plainly too unrealistic, because it leads to a divergent result in (9). 
[In this sense we might say that the fluctuation-dissipation theorem knows 
that particles have inertia.] A more reasonable Ansatz for $R$ must 
incorporate a ``relaxation time" or ``rise time" $\tau$ 
representing the time it takes the particle to accommodate itself 
to any sudden change $f(t)$. Such an Ansatz is, for example.,
\begin{equation}
R(t) = \mu (1 - e^{-t/r}) \Theta (t) , 
\end{equation}
which describes the classical motion of a particle subject to viscous 
friction. Without making so specific a choice, however, we will employ a 
cruder cutoff which should be adequate for times much greater than $\tau$:
\begin{equation}
R(t) = \mu\Theta (t - \tau) . 
\end{equation}
With this $R$, (9) can be integrated exactly [using the distributional 
identity, $P(\coth x) = d / dx\ln\sinh |x|$ to produce the following 
fundamental equation of quantum Brownian motion: 
\begin{equation}
\frac{1}{2} \langle \Delta x^{2} \rangle = \frac{\mu\hbar}{\pi} \ln 
\frac{\sqrt{\sinh\Omega |t - \tau |\sinh\Omega |t + \tau |}}{\sinh\Omega\tau}
 (\Delta t \ll \tau) ,
\end{equation}
where again $\Omega \equiv \pi / \beta\hbar$.\\

\noindent
Now strictly speaking, there is the inconsistency in 
our derivation of (12) that $C(t)$ is ill defined for a particle moving in an 
unbounded space, because $\langle x^{2} \rangle$ in equilibrium would be 
infinite, and (8) would therefore assume the indeterminate form 
$\langle \Delta x^{2} \rangle = \infty - \infty$. To overcome this 
problem, one could confine the particle in a very long ``box" (confining 
potential), it being intuitively clear that this could alter neither 
$\langle \Delta x^{2} \rangle$ nor $R(t)$ in the limit of an 
infinitely large such box.\\

\begin{center}
{\bf THREE LIMITING CASES OF\protect\\*[0.1cm]
THE GENERAL FORMULA (12)}\\
\end{center}

\noindent
The possible limiting cases of (12) are determined by the relative 
magnitudes of the three times $\tau, \beta\hbar$, and $\Delta t$, which 
we may call, respectively, the 
relaxation time, the ``quantum time," and the ``diffusion time." 
{\it A priori} there would be 
essentially 3!=6 distinct cases, but 
since we must have $\Delta t \gg \tau$ in order to 
apply (12), we will limit ourselves to only three of them. [It is 
nonetheless instructive to notice that (12) becomes self-contradictory for 
$\Delta t$ near $\tau$ since it then equates an intrinsically positive 
expression to a negative right-hand side. This implies that (11) 
could not be the exact 
response function for any system, even in principle. More generally, one can 
derive from (7) and the definition of $C(t)$, a positivity criterion which any 
putative response function must fulfill in order to be physically viable. We 
do not know how restrictive this criterion is in practice, but we have 
checked that the $R$ of (10) yields a mean-square displacement which is 
non-negative for all times, as one might have expected.] \\

\noindent
{\it Case 1:} $\beta\hbar \ll \tau \ll \Delta t$. This is the classical 
limit, 
and (12) reduces to the classical relation 
\begin{equation}
\frac{1}{2} \langle \Delta x^{2} \rangle = (\mu / \beta)\Delta t = \mu kT\Delta t ,
\end{equation}
or $\mu kT(\Delta t - \tau)$ if the leading correction is retained. \\

\noindent
{\it Case 2}: $\tau \ll \Delta t \ll \beta\hbar$. This is the extreme 
quantum limit, in which (time) (energy) $\ll \hbar$ for the 
time scale set by the diffusion time $\Delta t$ and the energy scale 
set by the thermal energy $kT$. In this limit (12) reduces to 
\begin{equation}
\frac{1}{2} \langle \Delta x^{2} \rangle = \frac{\mu\hbar}{\pi} \ln \frac{\Delta t}{\tau} ,
\end{equation}
or
\begin{eqnarray*}
\frac{\mu\hbar}{\pi} \ln \left[\left[\frac{\Delta t}{\tau}\right]^{2} - 1\right]^{1/2}
\end{eqnarray*}
if somewhat more precision is desired. It is noteworthy that the temperature 
has disappeared entirely from this expression (except insofar as it 
influences $\mu$ and $\tau$) suggesting a quantum Brownian motion due 
entirely to ``zero-point" fluctuations, which are present even at absolute 
zero. Indeed, the striking logarithmic dependence in (14) could also have been 
derived by first taking the zero-temperature limit of the 
fluctuation-dissipation theorem itself, and only then applying it to the 
$R$ of a diffusing particle.\\

\noindent
{\it Case 3}: $\tau \ll \beta\hbar \ll \Delta t$. Intermediate between 
cases 1 and 2, this situation might be described 
as one in which the relaxation occurs on quantum time-scales, although the 
diffusion time itself is already classically long. [A suggestive way to 
rewrite the inequality $\tau \ll \beta\hbar$ is as the relation between 
diffusion constants, $D_{\rm classical} \ll D_{\rm quantum}$, where 
$D_{\rm classical} = \mu\beta$, and $D_{\rm quamtum} = \hbar/m$, with $m$ 
taken from the ``viscous damping'' relation $\tau = \mu m$ envisaged in (10).]
In this case (12) reduces to
\begin{equation}
\frac{\langle \Delta x^{2} \rangle}{2} = \frac{\mu\Delta t}{\beta} + 
\frac{\mu\hbar}{\pi} \ln \frac{\beta\hbar}{2\pi\tau} ,
\end{equation}
which one can interpret as the result of two-stage spreading which follows 
the quantum law (14) up to the time $t_{Q} \equiv \beta\hbar / 2\pi$, and 
thereafter continues according to the classical law (13), with the second 
term in (15) remaining forever as a kind of residue of the quantum era. 
In order for this residue to be significant, we need 
$\mu\Delta t / \beta \leq (\mu\hbar / \pi)\ln (\beta\hbar / 2\pi\tau)$, or
\begin{eqnarray*}
\Delta t \leq \frac{\beta\hbar}{\pi} \ln \frac{\beta\hbar / \pi}{2\tau} ,
\end{eqnarray*}
which can occur nontrivially (i.e., without reducing to case 2) only if 
$\beta\hbar / \tau$ is exponentially big, so that $\ln (\beta\hbar / \tau) 
\geq \Delta t / \beta\hbar \gg 1$. Taking $m = \tau/\mu$ as earlier, 
this amounts to a requirement that the particle be extremely light: 
\begin{equation}
m \leq (\beta\hbar / \mu )e^{-\pi\Delta t / \beta\hbar} .\\
\end{equation}

\begin{center}
{\bf REMARKS AND NUMERICAL ESTIMATES}\\
\end{center}

\noindent
Equations (13), (14), and (15) are all special cases of the more general 
relation (12), which should be valid whenever $\tau \ll \Delta t$.
In other situations, or for 
more general response functions $R(t)$, one must refer back to (9) itself, 
from which the spreading can always be computed as 
long as $R(t)$ is known. A particularly interesting response function to treat 
would be (10), and another interesting case might be a particle moving in 
a superfluid.\\

\noindent
In the zero-temperature limit, i.e., in case 2 above, our formula (14) may 
be compared with a result of Ambegaokar,$^{4}$ who used a path-integral 
formalism, and assumed a linear coupling between the particle and an 
environment comprising an infinite collection of harmonic oscillators. He 
obtained an expression for the mean-square displacement of a Brownian 
particle in the quantum regime which corresponds to our result given in 
(14), if we make certain identifications. According to Ambegaokar (with a 
presumed misprint corrected), 
\begin{equation}
\langle (\Delta x^2) \rangle = [(h / \pi^{2}) / \gamma m]\ln |(t\sqrt{\omega_{c}\gamma)}| + {\rm const}.\;\; ,
\end{equation}
for $(1/\gamma ) < t < (\beta\hbar)$. Here, $\langle \Delta x^{2} \rangle$ is 
given by a density operator $\rho$ which reduces to a $\delta$ function at 
$t=0$, and $\omega_{c}$ defines an upper frequency cutoff beyond which the 
linear relationship between particle velocity and environmental friction 
breaks down. Also, judging from Eq. (5.3) of Ref. 4, it appears natural to 
identify $1/\gamma$ with our $\tau$, and therefore $1/m\gamma$ with our 
$\mu$. 
If we do so, and also equate $\omega_{c}$ to $\tau^{-1}$, then we recover 
(14) from (17) with the constant set to zero.\\

\noindent
In connection with (14) one can ask the following question: Classically, 
what kind of response function would lead to a logarithmic law of diffusion? 
If we take the $\hbar \rightarrow 0$ limit of (9), we find that the 
relevant response function should be proportional to $1/t$, which is 
physically impossible. This implies that the effect described by (14) is 
of purely quantum-mechanical origin.\\

\noindent
Finally, let us estimate the thresholds of time and temperature at which 
significant deviations from classical behavior should appear. In order to 
be in the ``pure quantum regime," we need $\Delta t \ll \beta\hbar$, which 
can also be written in the time-energy form, $kT\Delta t \ll \hbar$. Taking 
$T \sim 10^{-6}$ deg (cf Ref 6) and $\Delta t \sim 10^{-6}$ sec 
yields $kT\Delta t /\hbar \sim 0.1$, which ought to be well within the 
``pure quantum regime," 
meaning that (14) should apply if the relaxation time is short enough (and 
the ``reservoir" in thermal equilibrium). For higher temperatures or longer 
times, deviations of the sort described by (15) might be observable if 
$\tau$ is small enough and a condition like (16) is satisfied.\\

\begin{center}
{\bf ACKNOWLEDGMENTS}\\
\end{center}

\noindent
We would like to thank M. Cristina Marchetti for useful discussions and 
for making us aware of Ref. 3 and 4. Also R.D.S. would like to thank 
Lochlainn O'Raifeartaigh and Siddhartha Sen for their hospitality at the 
Dublin Institute for Advanced Studies, where parts of this paper were 
written. This research was partly supported by NSF grants No.PHY 9005790, 
No.INT 8814944, No.PHY 8918388, and No.DMR-87-17337. \\~\\

\noindent
$^{*}$ Permanent address: Department of Physics, Syracuse 
University, Syracuse N.Y. 13244-1130. \\~\\

\begin{itemize}
\item[]$^{1}$ A. O. Caldeira and A. J. Leggett, Physica A {\bf 121}, 587 
(1983). 
\item[]$^{2}$ H. Grabert 
and P. Talkner, Phys. Rev. Lett. {\bf 50}, 1335 (1983). 

\item[]$^{3}$ G. W. Ford, J. T. Lewis, 
and R. F. O'Connel, Phys. Rev. A {\bf 37}, 4419 (1988). 

\item[]$^{4}$ V. Ambegaokar, in {\it Frontiers of Nonequilibrium 
Statistical Physics} edited by Gerald T. Moore and Marian O. Scully 
(Plenum, New York, 1986), pp.231-239. 

\item[]$^{5}$ R. Balescu, {\it Equilibrium and Nonequilibrium, 
Statistical Mechanics}, (Wiley, New York, 1975), pp.663-669. 

\item[]$^{6}$ P. D. Lett, W. D. Phillips, S. L. Rolston, C. E. 
Tanner, R. N. Watts, and C. I. Westbrook. J. Opt. Soc. Am. B 6, 2084 (1989). 
\end{itemize}
\end{document}